\documentstyle[12pt]{article}

\newcommand{\inv}{^{-1}}
\def \del {\partial}

\newcommand{\r}{\rightarrow}

\newcommand{\ov}{\over}
\newcommand{\be}{\begin{equation}}
\newcommand{\ee}{\end{equation}}
\newcommand{\eel}[1]{\label{#1}\end{equation}}
\newcommand{\bea}{\begin{eqnarray}}
\newcommand{\eea}{\end{eqnarray}}
\newcommand{\eeal}[1]{\label{#1}\end{eqnarray}}
\newcommand{\baq}{\begin{equation}\begin{array}{rcl}}
\newcommand{\eaq}{\end{array}\end{equation}}
\newcommand{\eaql}[1]{\end{array}\label{#1}\end{equation}}
\newcommand{\beac}{\begin{equation}\begin{array}{rcl}}
\newcommand{\eeacn}[1]{\end{array}\label{#1}\end{equation}}
\newcommand{\ba}{\begin{array}}
\newcommand{\ea}{\end{array}}
\newcommand{\non}{\nonumber \\}

\newcommand{\al}{{\alpha'}}

\newcommand{\beq}{\begin{eqnarray}}
\newcommand{\eeq}{\end{eqnarray}}

\newcommand{\va}{\varphi}

\newcommand{\gym}{g_{YM}}

\begin{document}
\newcommand{\preprint}[1]{\begin{table}[t]  %%
           \begin{flushright}               %%
           \begin{large}{#1}\end{large}     %%
           \end{flushright}                 %%
           \end{table}}                     %%
\preprint{hep-th/9803103\\TAUP-2479-98\\Imperial/TP/97-98/29\\
}
\begin{center}
\LARGE{Supergravity Solutions  for  Branes
\\ Localized Within Branes}

\vspace{6mm}

\normalsize{N. Itzhaki$^1$,  A.A. Tseytlin$^2$  and S. Yankielowicz$^1$}

\vspace{6mm}

{\em ${}^{1}$ School of Physics and Astronomy, 
Tel Aviv University, Ramat Aviv, 69978,
 Israel\\
sanny, shimonya@post.tau.ac.il\\}

{\em ${}^{2}$ 
 Blackett Laboratory, 
  Imperial College, London SW7 2BZ, U.K.\\
tseytlin@ic.ac.uk\\}

\end{center}

%\vspace{5mm}

\begin{abstract}
We construct   supergravity solutions 
describing branes (D2-branes or  NS 5-branes or  waves) 
 localized  within D6-branes in the  region close to the
  core  of the D6-branes.
Other  similar   
string-theory and M-theory `near-core' localized solutions
  can be  found by applying U-duality
and/or lifting $D=10$ solutions to $D=11$.  
In particular, the   D2-branes localized on D6-branes is T-dual
to a special case of the 
 background describing  (D)strings localized on (D)5-branes
and  thus is also related to a localized intersection of 
M2-branes  and  M5-branes. 
D6+wave  configuration is  U-dual 
to a D0-brane localized on a Kaluza-Klein 5-brane
or to the fundamental string intersecting a D5-brane
with the point of intersection  localized on D5-brane.
\end{abstract}
%\newpage
\baselineskip 18pt

%%%%%%%%%%%%%%%%%%%%%%%%%%%%%%%
\section{Introduction}
%%%%%%%%%%%%%%%%%%%%%%%%%%%%%%%%%%%

In view of the fundamental role of soliton solutions of
supergravity there seems to be little need to justify 
 construction of new  such solutions.
In particular, recent progress in relating the large $N$ behavior
of certain gauge  theories
 to semi-classical supergravity 
(see, e.g., \cite{mal2}  and references there) 
serves as a  motivation for looking
for special solutions describing  regions close
to the cores of the branes. 
%In the large $N$ limit the quantum vacuum of a field
% theory is described by a {\em classical} supergravity
% solution.\footnote{The corresponding supergravity solutions 
%are mostly known only for theories with maximal supersymmetries.
%Solutions are also known
%for  some exotic theories with less supersymmetries.
%These solutions are obtained from the solutions with maximal 
%supersymetries by orbifolds identifications
%\cite{lim}.  HOW IMPORTANT IS TO KEEP THIS FOOTNOTE ?}
%,kac,mic,vafa}.}
Also, finding  supergravity solutions
representing branes  ending
on branes like 
 Hanany-Witten \cite{han}
  configuration (and other configurations in different 
dimensions and 
with less supersymmetries \cite{kut})  should be important.

%should shed  light  on the large $N$ behavior of these theories.
%The corresponding solutions  involve branes ending
%on branes.

General  solutions 
representing  localized  (as opposed to smeared \cite{pt,ttt}) 
intersections of branes
or branes ending on branes  have not yet been constructed
but some  progress have already being made.
The  solution  describing 
a  string localized on a  5-brane
was found (in an implicit form) in  \cite{TT,t}
 (see also   \cite{call}).
This is a special case of the solution corresponding to a  
string localized on 
 an intersection  \cite{khuri}
of  two 5-branes \cite{t,gaunt}. This latter solution and similar  composite 
solutions involving  several  branes 
 related    by U-duality  and lifting to $D=11$
(see also \cite{tat}) represent  only partially  localized intersections, 
while  in the case of, e.g., 
  the  Hanany-Witten configuration 
the localization should be in all relevant dimensions.

In this  note we present several {\it explicit}
 solutions  for branes which are 
{\it completely} 
{ localized} within other branes.
The 1/4 supersymmetric  solutions we find describe, however, 
only  the region  close to the core of the 
`bigger' brane. 
Our starting point  is the  $D=11$ 
solutions (M2-brane, M5-brane  and gravitational wave
 with 
$Z_N$ identifications in the transverse space)
which we shall reduce down to $D=10$ along a circular 
direction of  $S^3$ part of 
the transverse space.\footnote{This  
reduction  is a special 
case of the Hopf reduction discussed in detail in 
\cite{duf1} which appeared while this  paper  was in preparation
(the  motivation and interpretation  of the solutions found in 
\cite{duf1} is different from that of the present work).}
The M2-branes solution leads (Sec.2)  to $D=10$ 
background
representing D2-branes localized  on D6-branes 
in the region close to the D6-branes core
(or, equivalently,
 it corresponds to  the limit of very large charge of D6-brane). 
The M5-branes  and  waves  backgrounds   lead (Sec.3) to similar solutions 
describing the near-core region of D6-branes with, respectively,
NS 5-branes and
 waves localized on it. 

Other similar solutions can be  found by applying 
U-duality. 
In particular, T-dual  to our D6+D2  background  
 gives explicit analytic expression for  the 
`string localized  on 5-brane'  solution  (which was 
previously  found  only 
  in the general form   parameterized by 
 a function subject to a differential equation 
 \cite{t,TT})  in  the special  
case  when the 5-brane is smeared in one of the transverse 
directions  and in the region close to the core of the 5-brane
(Sec.4).  
The  system
of D1-branes localized within D5-branes  is
relevant for  the (4,4) two-dimensional 
 field theory description of certain
five dimensional black holes (see, e.g.,  \cite{mal3}).
Applying T-duality twice one  obtains a system of D0-branes 
localized within D4-branes  related to
the DLCQ of the superconformal (0,2) theory \cite{abs}.
Finally,  D6+D2 configuration  is also related by T-duality to a
D-instanton within D3-branes  system 
which corresponds to
a localized instanton in the ${\cal N}=4$ four-dimensional 
 theory.
Therefore, in the spirit of \cite{mal2}, the ${\cal N}=4$ 4d
 theory on a background of instantons  should be  dual to type IIB string 
 on a D3+D-instanton   background 
 T-dual to the D6+D2  one we  find here.

There are several possible generalizations. 
In particular, 
one may start with configurations of several intersecting 
M-branes and reduce along cycle of transverse 
$S^3$  obtaining near-core form of localized intersections 
involving  $n > 2$ different  D-branes.
One example is provided by the explicit
 near-core  form of solution 
describing  string localized on  the intersection of the two 5-branes
(Sec.4) which is related \cite{t} to  localized 
$5\cap 5\bot 2$  M-theory   and $5_{NS}\cap 5_{R}\bot 3$ type IIB 
theory 
configurations.

%%%%%%%%%%%%%%%%%%%%%%%%%%%%%%%%%%%%%%%%
\section{D2-brane within D6-brane}
%%%%%%%%%%%%%%%%%%%%%%%%%%%%%%%%%%%%%%%
%\subsection{

We would like to find  a  supergravity solution 
representing  $n_2$
D2-branes localized within a collection of $N$
D6-brane in the decoupling  (or `near-core') limit.  From the point of view 
of the field theory  on the 
D6-brane world-volume,  it should  correspond
 to   localized 
2-branes `instantons'.
%which break half of the supersymmetries.
The  simplification  of  considering D6-branes in this limit
is that their  M-theory  counterpart
(KK monopole  background  \cite{towns})
becomes  an ALE space
 with an $A_{N-1}$ singularity \cite{gh} times a  7-dimensional 
Minkowski space $M^{(6,1)}$
 \cite{sen,sei}.
This  is just  a
Minkowski space $M^{(10,1)}$ with $Z_N$ identifications.
Suppose we  introduce  $n_2$ M2-branes along 2+1 directions 
in $M^{(6,1)}$.
Such M2-branes  are invariant under
 the $Z_N$ identifications.
 Therefore,  the corresponding 
eleven dimensional  background 
 will be given by  the  M2-brane solution \cite{duff}
with the $Z_N$ 
identifications in  transverse
 space.
%\footnote{An alternative way 
% is to consider
%the case of $N=1$, which is simply an eleven dimensional
% Minkowski space and to observe that the Dirac string
% singularity,  found  in the  $D=10$ solution, can be removed
%not only for $N=1$ but for any integer $N$.}   
%The solution is invariant under the identifications and 
%hence it
It can then  be reduced to ten dimensions to obtain
a type IIA solution  
which, remarkably, can be interpreted as representing 
the near-core region of a configuration  of  D2-branes
 localized within a 
collection of  D6-branes.

Before we  turn to  the D6+D2 case let us first
 recall how one  can obtain the near-core region of 
type IIA solution 
for  a collection of $N$ D6-branes 
upon dimensional reduction
of the $A_{N-1}$  space \cite{juan}.
The metric which corresponds to $M^{(6,1)}$ times 
an $A_{N-1}$
 singularity   is 
\be\label{An}
ds^2_{11} =dx^2_{||}+ d\rho^2 + \rho^2 (d\tilde \theta^2 +
  \sin^2\tilde \theta d\tilde\varphi^2 +
\cos^2\tilde \theta d\tilde \phi^2 ),
\ee 
where $dx^2_{||}=-dt^2+dx_1^2+...+dx_6^2$,\ \ 
$\rho^2=x_7^2+...+x_{10}^2$, \ \  
 $0\leq \tilde \theta \leq \pi/2$ and $0 \leq \tilde
 \varphi,
\tilde \phi \leq 2\pi$ with the $Z_N$
 identification $(\tilde \varphi, 
\tilde \phi) \sim (\tilde \varphi, \tilde \phi) + ( 2\pi/N,
 2\pi/N)$.
Defining the new variables 
\be\label{l}
U=\frac{\rho^2}{2N l_p^3},~~~ \theta=2\tilde
 \theta,~~~~\varphi=\tilde \varphi-\tilde \phi,~~~~
\phi=N\tilde \phi, 
\ee
 we obtain the metric
\be
ds^2_{11} =dx^2_{||}+\frac{l_p^3N}{2 U}dU^2+{ l_p^3NU \over 2}
 (d\theta^2+
\sin^2 \theta d\varphi^2)
+\frac{2 U l_p^3 }{N}[d\phi+ { N \over 2 } (\cos\theta -1)
 d\varphi ]^2 ,
\ee
where $\phi$ has standard 
 period $\phi \sim 
\phi + 2 \pi$. 
This metric has a Killing vector along the $\phi $ 
direction and hence it can be reduce to ten dimensions along
$x_{11} \equiv R_{11} \phi $.
Using the relation between  the eleven dimensional  metric
 and the ten dimensional type IIA string metric, dilaton and
the 
gauge field, 
\be\label{1011}
ds^2_{11}=e^{4\phi /3 }(dx_{11}+A_{\mu}dx^{\mu})^2+
e^{-2\phi /3 }ds^2_{10},
\ee 
we obtain 
\beq\label{solD6}
&& ds^2_{10} =\al\left[ \frac{(2\pi)^2 }{\gym}\sqrt{\frac{2 U}
{ N}}dx^2_{||}+
{\gym \over (2\pi)^2 }
\sqrt{\frac{N}{2 U}}dU^2+{\gym \over (2\pi)^2 \sqrt{2}}
\sqrt{N}U^{3/2}d\Omega^2_2\right], \non
&& e^{\phi}= { g^2_{YM} \over 2 \pi} 
\left( \frac{2U}{\gym^2N}\right) ^{3/4},\\
&& A_{\mu}dx^{\mu}=\frac{N}{2} (\cos\theta -1)d\va,\nonumber
\eeq
where $d\Omega^2_2=d\theta^2+\sin^2\theta d\va^2$. 
This is the IIA supergravity solution  representing  a collection of
D6-branes \cite{sh}
 in the decoupling limit \cite{mal}
\beq\label{limD6}
&& U=\frac{|x|}{\al}=\mbox{fixed},\;\;~~~~
\gym^2=(2\pi)^4 l_p^3=(2\pi)^4 g_s\al^{3/2}=
\mbox{fixed},\;\;\;~~~~
\al\r 0.
\eeq
%where $\gym^2=$.

Let us now  consider 
a  configuration of M2-branes 
stretched along  $x_1,x_2$  directions.
For simplicity we shall put all of them at the origin
in the transverse space 
(it is straightforward to generalize the solution to the 
case where the two-branes are separated along $x_3,...,x_6$).
The required eleven dimensional solution is 
obvious:
it is essentially the same as in 
\cite{duff}
 but with the above  $Z_{N}$ identification
of angles of $S^3$ 
(the relevant harmonic function depends 
 only on the radial coordinate $\rho$
via $\rho^2 + x_i x_i$   and not on $\tilde\theta, \tilde\phi$ and 
$\tilde\va$).
The  metric  is,  therefore, 
%(we consider only the metric for simplicity),
%\beq
\[
%&&
 ds^2_{11} =f_2^{-2/3}(-dt^2+dx_1^2+dx_2^2)\non
%&&
\]
\[+ \  f_2^{1/3}\left[ dx_3^2+...+dx_6^2+d\rho^2+\rho^2(d\tilde 
\theta^2 +  \sin^2\tilde \theta d\tilde\varphi^2 +
\cos^2\tilde \theta d\tilde \phi^2 ) \right],
\]
%\eeq
where 
\be\label{m}
f_2=1+\frac{2^5\pi^2 n_2l_p^6}{\hat r^6},~~~~~~~~ \hat r^2=x_3^2+
...+x^2_6+\rho^2,~~~~~~\rho^2=x_7^2+...+x_{10}^2, 
\ee  
and the 3-rank tensor has the standard form  $C_{012}=f_2^{-1}$. 
Changing the coordinates  as in (\ref{l}) we get
\beq
&& ds^2_{11} =f_2^{-2/3}(-dt^2+dx_1^2+dx_2^2)+ f_2^{1/3}\left(
 dx_3^2+...+dx_6^2+\frac{l_p^3N}{2 U}dU^2 \right.\non
 && \left.
+ \ { l_p^3NU \over 2}
 (d\theta^2+
\sin^2 \theta d\varphi^2)
+\frac{2 U l_p^3 }{N}[d\phi+ { N \over 2 } (\cos\theta -1)
 d\varphi ]^2 \right), 
\eeq
where now
\be
f_2=1+\frac{2^5\pi^2 n_2l_p^6}{(x_3^2+
...+x^2_6+2Nl_p^3U)^3} \ . 
\ee
As  above,   we can reduce the solution to ten dimensions
along the isometric 
$\phi$ direction.  The result is the 
 type IIA solution  describing D2-branes localized 
within D6-branes (in  the  region  near the core of 
 D6-branes) 
\beq\label{o}
&&
 ds_{10}^2=\al\left[ f_2^{-1/2} h^{-1/2}_6
(-dt^2+dx_1^2+dx_2^2)+
f_2^{1/2} h^{-1/2}_6(dx_3^2+...+dx_6^2) 
\right.\non && \left.
+ f_2^{1/2}h_6^{1/2}(dU^2+U^2 d\Omega^2_2)\right] , \non
&& e^{\phi} ={ g^2_{YM} \over (2 \pi)^4} 
 f_2^{1/4} h_6^{-3/4} ,\\
&& A_{\mu}dx^{\mu}=\frac{N}{2} (\cos\theta -1)d\va,
\nonumber 
\eeq
where
\be\label{i}
h_6=\frac{\gym^2 N}{2 (2\pi)^4 U } , 
\ee
and $C_{012}= f^{-1}_2$. 
Note that the presence of the D2-branes does not modify
the expression for 
gauge field $A_\mu$ and hence the Dirac string singularity can
 be removed for any integer $N$ 
%(as expected from the 
%discussion at the beginning of  this  section).  

This background has the same `harmonic function rule' form 
as  the standard 
1/4 supersymmetric  asymptotically flat 
D6+D2  bound state solution 
\cite{ttt}
for which  the positions of 
 D2-branes are smeared within
the D6-brane directions  so that  $f_2$ and $h_6$ are given by 
\beq
f_2 =1  + {\bar Q_2\over r}, \ \ \ \ \ \ \  h_6 = \al^2 f_6 \  ,  \ \ \ \ \ \
 f_6  =  1    + {Q_6\over r}
\ 
,\ \  \ \   \ \ \  Q_6 = {1\over 2} N g_s \al^{1/2} \ . 
\label{old} 
\eeq
Here instead  we find 
\beq
 f_2 = 1 + {Q_2\over (x^2_3 +...+ x^2_6 + {  4}  Q_6 r)^3} ,
 \ \ \ \ \ \ \ \ \ \ \ \ \ \
h_6= \al^2 {Q_6\over r}, \ \ \ \ \ \  \ \ r =  {U/ \al} \ , 
\label{new} 
\eeq
i.e. 
we are restricted to the region  close to  the D6-brane core
(or, equivalently, consider the decoupling limit  (\ref{limD6}))  but   the 
D2-branes are  completely localized  within D6-branes
(`averaging' over $x_3,...,x_6$  leads  back to the 
near-core region of the smeared solution). 
Surprisingly,  the   D2-brane function 
$f_2$ in  (\ref{new}) `remembers' its 
$D=11$ membrane origin:    its
power of decay with distance is $1/x^6$ compared to $1/x^5$
for the usual D2-brane in  ten dimensions.
This behavior  is characteristic of a string solution, and indeed
the above D6+D2 configuration is T-dual to a D-string localized on 
D5-brane (see Sec.4).
Note that the $1/x^6$ is the behavior at the near core region 
(where our solution is valid).
Thus in the full solution the power of decay with distance
should interpolates between  $1/x^6$ at the  near core region 
 and $1/x^5$ far from the horizon.

To summarize,  the solution we 
have found 
applies   only in the decoupling limit.
It  should be  a `near-core' approximation to
a more general  solution, yet to be found,
 describing D2-branes localized within D6-branes
and  having also  the  asymptotically flat region
(or, equivalently, to the solution with  finite $\al, g_s$ and $r$).
Trying to add a constant to $h_6$ in  (\ref{new})
does not,  however,  lead to a simple analytic 
expression for the background (cf. Sec.4).

%In the smeared solution the extension of  the
% ``near-core'' solution to the large distances region is
%obtained by adding a constant to $h$ (which in our case 
%should be $\sim\al^2$ to get the standard result).

%We do not think that this is the case with the localized 
%solution eq.(\ref{o}) because in our derivation 
% the subtraction of the $1$ in the harmonic function 
%of the Taub-NUT solution plays a crucial role. 

%%%%%%%%%%%%%%%%%%%%%%%%%%%%%%%%%%%%%%%%%
\section{NS5-brane  and wave within D6-brane}
%%%%%%%%%%%%%%%%%%%%%%%%%%%%%%%%%%%%%%%%%%%%%%%%%%%%%%
To find  a  solution describing 
 $n_5$ NS5-branes  within a 
collection of $N$ D6-branes 
we  follow the logic of the previous section
  starting now with the  M5-brane solution \cite{guv}
with the  $Z_N$ identification in the transverse 3-sphere.
Its metric is 
\[
%\beq
%&&
 ds^2_{11} =f_5^{-1/3}(-dt^2+dx_1^2+...+dx_5^2)+
f_5^{2/3}\left[ dx_6^2+d\rho^2
%\right.\non &&\left.
 + \rho^2(d\tilde 
\theta^2 +  \sin^2\tilde \theta d\tilde\varphi^2 +
\cos^2\tilde \theta d\tilde \phi^2 ) \right] ,
\]
%\eeq
where
\be
f_5=1+\frac{\pi n_5 l_p^3}{\hat r^3},~~~~~~~~~~~~~~~~\hat r^2=x_6^2+\rho^2, 
\ee
and $\rho$  was  defined in (\ref{m}).
The change of variables  (\ref{l}) gives
\beq
&& ds^2_{11} =f_5^{-1/3}(-dt^2+dx_1^2+...+dx_5^2)+
 f_5^{2/3}\left(dx_6^2+\frac{l_p^3N}{2 U}dU^2\right.\non
 && \left.
+\  { l_p^3NU \over 2}
 (d\theta^2+
\sin^2 \theta d\varphi^2)
+\frac{2 U l_p^3 }{N}[d\phi+ { N \over 2 } (\cos\theta -1)
 d\varphi ]^2 \right), 
\eeq
where 
\be
f_5=1+\frac{\pi n_5 l_p^3}{(x_6^2+2Nl_p^3U)^{3/2}},
\ee
Reducing to  ten dimensions along $\phi$ 
we obtain the type IIa solution of
$n_5$ NS-fivebranes localized within $N$ D6-branes,\footnote{This 
is to be compared with the solution describing 
delocalised superposition of D6-brane  and NS 5-brane 
which is  a reduction of  the 
`smeared'  superposition of M5-brane and 
KK monopole given in \cite{ttt}. }  
\beq
&& ds_{10}^2=\al\left[  h^{-1/2}_6
(-dt^2+dx_1^2+...+dx_5^2)+
%\right.\non && \left.
f_5 h^{-1/2}_6 dx_6^2+
f_5 h^{1/2}_6(dU^2+U^2 d\Omega^2_2)\right], \non
&& e^{\phi} ={ g^2_{YM} \over (2 \pi)^4} 
f_5^{1/2} h^{-3/4}_6,\\
&& A_{\mu}dx^{\mu}=\frac{N}{2} (\cos\theta -1)d\va,
\nonumber 
\eeq
where $h_6$  was defined in eq.(\ref{i}). 
Reducing instead  along $x_5$ we find a superposition of 
D4-brane and the near-core  region of  KK monopole. 
T-duality in $x_6$ and $\phi$ then leads back 
to D6+ NS5 solution.

Next we turn to waves localized within D6-branes.
Starting with  the  metric of the gravitational wave
in $D=11$
$$
ds^2_{11} = -dt^2+ dx_1^2 + f_0 (dt-dx_1)^2
 + dx_2^2+...+ dx_6^2+d\rho^2
 + \rho^2(d\tilde 
\theta^2 +  \sin^2\tilde \theta d\tilde\varphi^2 +
\cos^2\tilde \theta d\tilde \phi^2 ) ,
$$
$$ f_0 = {Q_0 \over \hat r^7} \ , \ \ \ \ \ \ 
\hat r^2 = x^2_2 + ...+x^2_6 + \rho^2 \ , 
$$
and reducing it  along $\phi$ as above
we find the  background representing a wave localized
 on D6-brane  with the metric 
\beq
&& ds_{10}^2=\al\left(  h^{-1/2}_6
[-dt^2+ dx_1^2 + f_0 (dt-dx_1)^2
 + dx_2^2+...+ dx_6^2]
%\right.\non && \left.
+  h^{1/2}_6(dU^2+U^2 d\Omega^2_2)\right). \non
\label{kkk}
\eeq
T-duality along the   wave ($x_1$) 
 direction  gives a configuration of 
D5-brane (smeared along $x_1$) 
 intersected by 
a fundamental  string   with  the  intersection point localized on D5. 
S-dual to this is NS5-brane  intersected by a  D-string.
 T-duality along D-string direction $x_1$ 
gives then a near-core 
configuration 
 of Kaluza-Klein  5-brane 
(which is T-dual to NS 5-brane  smeared in one transverse
direction with T-duality applied in that direction) 
 with D0-brane  completely 
 localized on it in all internal 
coordinates.
But this is just the background 
 which is obtained  directly 
by reducing (\ref{kkk}) 
along  the $x_1$ direction. Again, as usual, the 
reductions of $D=11$ solutions 
 along different isometric directions 
give U-dual $D=10$ solutions.

%%%%%%%%%%%%%%%%%%%%%%%%%%%%%%%%%%%%%%%%%
\section{String localized on 5-brane}
%%%%%%%%%%%%%%%%%%%%%%%%%%%%%%%%%%%%%%%%%
Applying  T-duality to the localized  D6+D2 solution 
 (\ref{o})  along $x_2$ direction one expects to find
 the 
D5+D1 solution   or  S-dual NS5+NS1 solution 
describing  a  fundamental string  \cite{dab}
localized on a NS5-brane \cite{cal}. 
 Let us show that, indeed, 
 the  background  T-dual to   (\ref{o}) 
is a {\it special case}  of the general 5+1 solution constructed in 
\cite{TT,t}.
The exact string  background   describing
NS 5+1 configuration  is represented by  the 
conformal sigma-model   with the following metric, 
2-form and dilaton couplings ($m,n=1,2,3,4$) 
\beq
ds^2_{10} = f_1^{-1} (x,y) (-dt^2 + dz^2) +  dy_n dy_n  + f_5(x) dx_m dx_m
 \ , 
\label{soo}
\eeq
 $$
dB= df\inv_1 \wedge dt\wedge dz  + *df_5  \ , \ \ \ 
e^{2\phi} = f_1\inv   f_5   \ . $$
Here  $(z,y_n)$ are the internal dimensions of the 5-brane
($x_1,x_3, ..., x_6$ in  (\ref{o})), 
$x_n$ are the dimensions transverse to the 5-brane  ($x_2$, 
$ U, \theta, \varphi$  in  (\ref{o})) and  
 $f_5(x)$ is the harmonic function 
($\del^m \del_m f_5 \equiv \del^2_x f_5=0$)
 which defines  the position of the 5-brane(s).
The string function $f_1(x,y)$  must 
 satisfy the condition
(Laplace equation in the curved transverse space)
\beq
 [  \del^2_x +  f_5(x)  \del^2_y] f_1 (x,y) =0 \  . 
\label{eqq}
\eeq
Remarkably, this equation admits a simple analytic solution in the 
special case 
when  (i) the 5-brane is smeared in one of the transverse dimensions
(e.g., $x_2$) 
so that $f_5= 1 + Q_5/r , \ \   r^2 = x_1^2 + x_3^2 + x^2_4, $
and\ \  (ii) one considers only the near-throat region of the   5-brane 
where
$f_5 \to  Q_5/r$.  Assuming that $f_1$ depends  only 
on  the radial coordinates $r$ and $v, \ v^2\equiv  y_m y_m$, 
we can put (\ref{eqq}) in the form
\beq
 [   r^{-1} \del_r (r^2 \del_r)  +  
    Q_5     v^{-3}  \del_v (v^3 \del_v)        ] f_1 (r,v) =0 \  .
\label{eqeq}
\eeq
In terms of   the variable $u= 2 \sqrt{ Q_5 r}$ it becomes 
\beq
 [   u^{-3} \del_u (u^3 \del_u)  +  
        v^{-3}  \del_v (v^3 \del_v)        ] f_1 (u,v) =0 \  ,
\label{qeiq}
\eeq
i.e.  is formally the same as the  radial part of the 
Laplace equation  in {\it flat} 
  8 dimensions (with $u^2 +v^2$ as the  total distance).
This equation is  solved, in particular,  by 
\beq
f_1 = 1 + {Q_1 \ov (v^2 +u^2)^3} = 1 + {Q_1 \ov (y^2_1 + ...+y^2_4  +  4Q_5 r )^3}
\ , 
\label{qeq}
\eeq
which has, indeed,  the same form as $f_2$ in  (\ref{new}). 
More general solutions are found  by separating the string cores  in 
the 5-brane 
$y_n$ directions. This is in agreement with the possibility to choose a 
more general 
M2-brane harmonic function $f_2$ in (9) (e.g.  a sum 
of  several terms with different centers in $x_3,...,x_6$)
and  allows, in particular, to   construct the solution smeared 
in these internal directions, thus  returning back to $f_2$ in (\ref{old}). 

One  straightforward  generalization 
is obtained by  replacing 
 the  product of the flat  4-space $y_m$ 
 and curved 5-brane 4-space $x_n$ factors  in (\ref{soo})
 by the direct product of the two 5-brane factors  \cite{khuri}. 
The resulting  exact conformal 
background  describing  a 
string localized on  the 
intersection of the  two 5-branes 
 is  \cite{TT,t} 
\beq
ds^2_{10} = f_1\inv (x,y) (-dt^2 + dz^2) + f'_5(y)
 dy_n dy_n  + f_5(x) dx_m dx_m \ , 
\eeq
$$
dB= df\inv_1 \wedge dt\wedge dz  + *df_5 +  *df'_5 \ , \ \ \  \ \ 
e^{2\phi} = f_1\inv   f_5  f'_5   \ , $$
where  the harmonic functions 
$f_5(x)$ and $f_5'(y)$  ($ \del^2_x f_5 =0, \ \del^2_y f_5'=0$)
define the positions of the two 5-branes $(z,y_n)$ and $(z,x_m)$
 and the string function  $f_1(x,y)$ satisfies
\beq
 [  f_5'(y) \del^2_x +  f_5(x)  \del^2_y] f_1 (x,y) =0 \  . 
\label{uuu}
\eeq
Assuming that the two 5-branes are smeared in one of the 
relative transverse directions (e.g., $x_2$ and $y_2$) 
and considering the near-core region
where 
$$
f_5 = {Q_5\ov r},  \ \ \ \ \ \ \ 
f'_5 = {Q_5'\ov r'},  \ \ \ \ \   r^2 =x_1^2 + x^2_3 + x^2_4, \ \ \
 r'^2 =y_1^2 + y^2_3 + y^2_4,
$$
we find that (\ref{uuu})
is solved by 
$$
f_1 = 1  +    {Q_1 \ov ( Q_5 r +  Q'_5 r' )^3}
\ . 
$$
This gives  a simple  explicit  solution 
describing a string localized on an  intersection of the two 
5-branes. Closely related $D=10$ and $D=11$ solutions may be found 
by using U-duality and lifts to $D=11$ as in 
\cite{t,gaunt}.

%%%%%%%%%%%%%%%%%%%%%%%%%%%%%%%%%%%%%%%%%%%%%%%%%%%%%

To study   some  consequences of   localization 
of one brane on another  one may  consider  the action of a 
D5+D1 probe moving in the background produced by localized 
D5+D1 source. Equivalent T-dual system is D4+D0 probe 
moving in D4+D0 source background. 
Representing the D4+D0 probe
by the standard Born-Indeld action with a self-dual gauge field 
background  and ignoring the dependence on internal dimensions 
of D4-brane probe  one  finds  
the following expression for the probe action (see \cite{chep})
   $$
 I_{4+ 0}    =-  T_0  \int dt\ 
  \bigg(\  n_4 + n_0  
$$
$$ + \  [n_4  f_0^{-1} (x,y)  + n_0 f_4^{-1}(x) ] 
[ \sqrt {1-   f_0(x,y)  \dot y_m \dot y_m - 
 f_0(x,y)  f_4 (x)  \dot x_n \dot x_n }
    -1]
\bigg)  \  . $$
Here $f_0, f_4$ are the same as $f_1,f_5$ in   (\ref{eqq}) 
or as $f_2,f_6$ in  (\ref{new}) 
(originating from D6+D2, the D4+D0 background is smeared in the two transverse directions)
and $n_0,n_4$ are the charges of the D4+D0 probe. 
It is  obvious that  the motion of the probe  in  both the 
transverse   ($x_n$) 
and parallel ($y_m$)   directions to the  D4+D0  source 
depends on  the detailed  structure of $f_0(x,y)$.

%%%%%%%%%%%%%%%%%%%%%%%%%%%%%%%%%
{\bf Acknowledgments}
%%%%%%%%%%%%%%%%%%%%%%%%%%%%%%%%%%%%%%%%%%

The  work of N.I. and S.Y.  was  supported in part by the US-Israel
 Binational Science Foundation, by GIF --  
  the German-Israeli Foundation for Scientific Research, 
and by the  Israel Science Foundation.
The work of A.A.T.  was  supported  by
 PPARC and  the European
Commission TMR programme grant ERBFMRX-CT96-0045.
%%%%%%%%%%%%%%%%%%%%%%%%%%%%%%%%%%%%%%%%%%%%%%%%%%%%%%%%%%%

\end{document}